\documentclass[3p,times,procedia]{elsarticle}
\usepackage{nupha_ecrc}

\volume{00}

\firstpage{1}

\journalname{Nuclear Physics A}

\runauth{}

\jid{nupha}

\jnltitlelogo{Nuclear Physics A}

\def\snn{\mbox{$\sqrt{s_{_{\rm NN}}}$}}

\newcommand{ \be }{\begin{eqnarray}}
\newcommand{ \ee }{\end{eqnarray}}

\newcommand{ \bp }{{\bf p}}

\newcommand{ \bq }{{\bf q}}

\newcommand{ \eps }{\varepsilon}

\newcommand{\kt}{k_{\rm{T}}}

\newcommand{\rside}{R_{\rm side}}
\newcommand{\rout}{R_{\rm out}}
\newcommand{\rlong}{R_{\rm long}}
\newcommand{\ros}{R_{\rm os}}
\newcommand{\rol}{R_{\rm ol}}
\newcommand{\rsl}{R_{\rm sl}}
\newcommand{\qside}{q_{\rm side}}
\newcommand{\qout}{q_{\rm out}}
\newcommand{\qlong}{q_{\rm long}}
\newcommand{\rsidez}{R_{{\rm side},0}}

\newcommand{\rsidetwo}{R_{{\rm side},2}}

 \newcommand{\dphi}{\Delta\varphi}

\usepackage{amssymb}

\usepackage[figuresright]{rotating}

\begin{document}

\begin{frontmatter}



\dochead{XXVIth International Conference on Ultrarelativistic Nucleus-Nucleus Collisions\\ (Quark Matter 2017)}

\title{Azimuthally differential pion femtoscopy relative to the second and third harmonic in Pb--Pb 2.76 TeV collision from ALICE}


\author{Mohammad Saleh (for the ALICE Collaboration)}

\address{Wayne State University, 42 W. Warren Ave, Detroit, MI 48202}

\begin{abstract}
  Azimuthally differential femtoscopic measurements, being sensitive to spatio-temporal characteristics
  of the source as well as to the collective velocity fields at freeze-out, provide very
  important information on the nature and dynamics of the system evolution. While the HBT
  radii modulations relative to the second harmonic event plane reflect mostly the spatial
  geometry of the source, the third harmonic results are mostly defined by the velocity fields~\cite{Voloshin:2011mg}.
  Radii variations with respect to the third harmonic event plane unambiguously signal a collective
  expansion and anisotropy in the flow fields. Event shape engineering (ESE) is a technique proposed to select events corresponding
  to a particular shape. Azimuthally differential HBT combined with ESE allows for a
  detailed analysis of the relation between initial geometry, anisotropic flow and the deformation
  of source shape. We present azimuthally differential pion femtoscopy with respect to second
  and third harmonic event planes as a function of the pion transverse momentum for different
  collision centralities in Pb--Pb collisions at \snn~=~2.76 TeV.
  All these results are compared to existing models. The effects of the selection
  of the events with high elliptic or triangular flow are also presented.
\end{abstract}


\end{frontmatter}


\section{Introduction}
\label{Intro}
The correlation of two identical particles
at small relative momentum, commonly known as intensity, or Hanbury Brown-Twiss (HBT), interferometry, is an effective
tool to study the space-time structure of the emitting source in relativistic heavy-ion collisions~\cite{Bertsch:1988db}. Due to the position-momentum correlations
in particle emission, the so-called HBT radii become sensitive to the collective velocity fields, from which information about the
dynamics of the system evolution can be extracted.
Azimuthally differential femtoscopic measurements can be performed relative to the direction of different
harmonic flow planes~\cite{Voloshin:1995mc}. In particular, measurements of the HBT radii relative to the second harmonic flow provide information
on the final shape of the system, which is expected to become more spherical compared to the initial state due to stronger
in-plane expansion. In contrast, the azimuthal dependence of the HBT radii relative to the third harmonic flow plane
can originate only from the the anisotropies in collective flow gradients, and the observation of any HBT radii azimuthal dependence
will unambiguously signals a collective expansion and anisotropy in the flow fields~\cite{Voloshin:2011mg}. Event shape engineering (ESE) is a relatively
new technique which selects on the magnitude of the flow vector~\cite{Schukraft:2012ah}. This selection can provide more control on the
initial shape of the source.
 \section{Experimental analysis and results}
 \label{exp}
The data  Pb--Pb data used in this analysis were collected by ALICE in 2011. The Time
Projection Chamber (TPC) was used to
reconstruct the tracks in the pseudorapidity range $|\eta|<0.8$ as well as to identify pions via the specific ionization energy
loss d$E$/d$x$ associated with each track. In addition to the TPC, the Time-Of-Flight (TOF) detector was used for identifying pions with transverse
momentum $p_{\mathrm T}$~$>$~0.5 GeV/c. The second and third harmonic event plane angles, $\Psi_{2}$ and $\Psi_{3}$, were determined using TPC tracks.
\newline
The correlation function $C({\bq})$ was calculated as the ratio of $A({\bq})$ and $B({\bq})$, where
 $\bq=\bp_1-\bp_2$ is the relative 3-momentum of two pions,
$A(\bq)$ is the same-event distribution of particle pairs, and
$B(\bq)$ is the background distribution of uncorrelated particle
pairs obtained using mixed events technique.
The Bertsch-Pratt~\cite{Pratt:1986cc} out--side--long
coordinate system was used~\cite{Pratt:1986cc}.
The correlation function is fitted to
\begin{eqnarray}
 C({\bq},\dphi)=N[(1-\lambda)+\lambda K({\bq})(1+G({\bq},\dphi))],
\end{eqnarray}
where $N$ is the normalization factor, $\lambda$ is the chaoticity, and $\dphi=
\mathrm{\varphi_{pair}}-\Psi_{2(3)}$. The function $G({\bq},\dphi)$
describes the Bose-Einstein correlations and $K({\bq})$ is the Coulomb
correction. In this analysis the Gaussian form of
$G({\bq},\dphi)$ was used~\cite{guassain}:
\begin{eqnarray}
 G(\bq,\dphi)=\exp
\left[
-\qout^{2} \rout^{2}(\dphi)-\qside^{2} \rside^{2}(\dphi)
-\qlong^{2} \rlong^{2}(\dphi)-2\qout \qside \ros^{2}(\dphi) \right.
\nonumber
\\
\left.
-2\qside \qlong \rsl^{2}(\dphi)-2\qout \qlong \rol^{2}(\dphi)
\right],
\end{eqnarray}
where the parameters $\rout$, $\rside$, and $\rlong$ are traditionally
called HBT radii in the {\it out}, {\it side}, and {\it long}
directions.  The cross-terms $\ros^{2}$, $\rsl^{2}$, and $\rol^{2}$
describe the correlation in the {\it out}-{\it side}, {\it side}-{\it
  long}, and {\it out}-{\it long}~directions, respectively.
  \begin{eqnarray} \label{eq:radii_osc}
  R^{2}_{\mu}(\dphi)=R^{2}_{\mu,0}+2R^{2}_{\mu,2}\cos(2\dphi),~\ros^{2}(\dphi)=R^{2}_{{\rm os},0}+2 R^{2}_{{\rm os},2}\sin(2\dphi),
 \end{eqnarray}
  where~$\mu={\rm out,side,long,sl, and~ol}$. Fitting the radii's azimuthal dependence with the functional
  form of Eq.\ref{eq:radii_osc} allows us to extract the average radii and the amplitude of oscillations~\cite{Adamova:2017opl} for the second harmonic results.
\begin{figure}[!ht]
\hspace*{34mm}
\includegraphics[width=8cm,keepaspectratio]{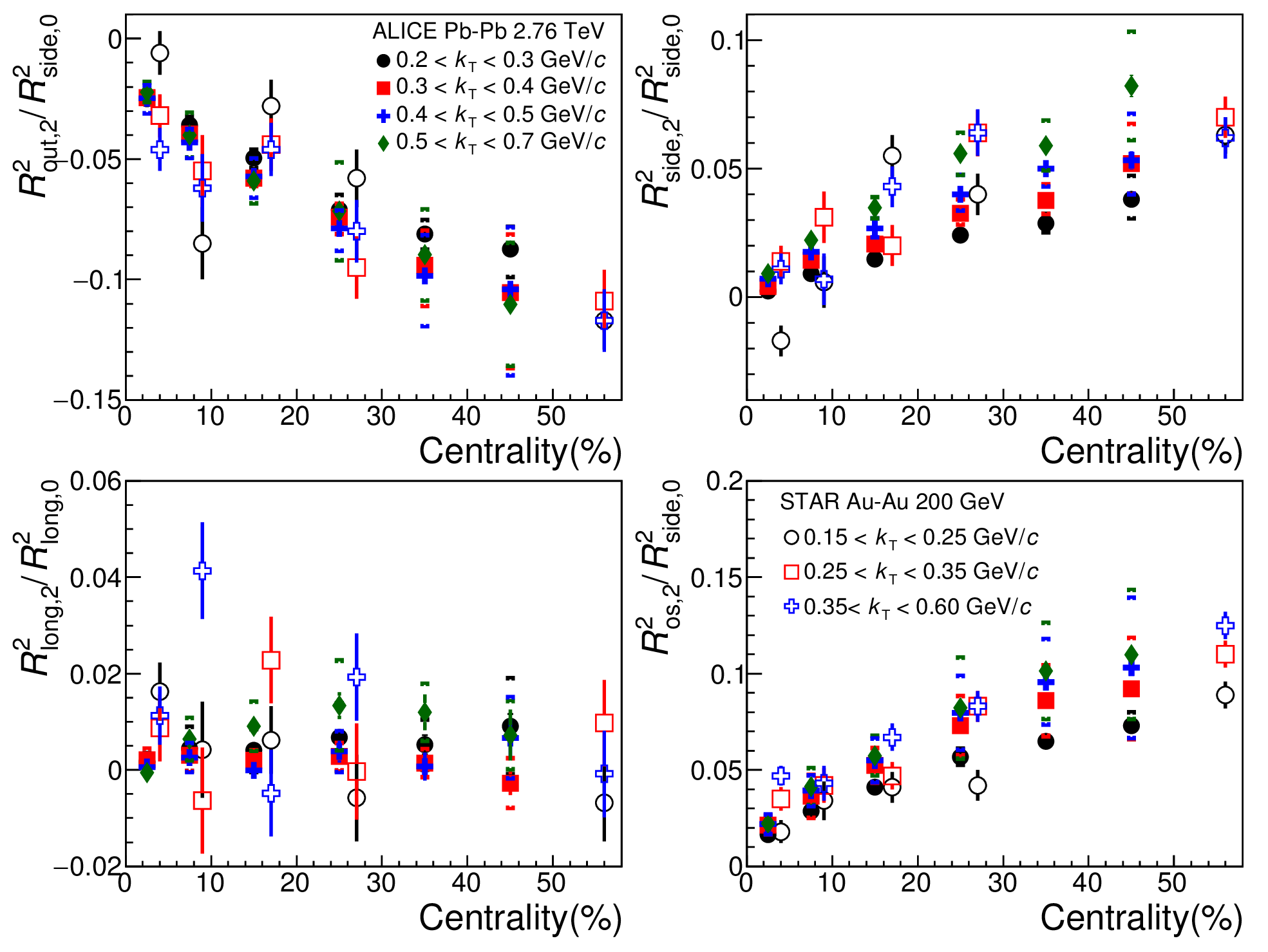}
  \caption{ Amplitudes of the relative radius oscillations~$R^{2}_{\rm
      out,2}/R^{2}_{\rm side,0}$,~$R^{2}_{\rm side,2}/R^{2}_{\rm
      side,0}$,~$R^{2}_{\rm long,2}/R^{2}_{\rm long,0}$,
    and~$R^{2}_{\rm os,2}/R^{2}_{\rm side,0}$ versus centrality for
    four $\kt$ ranges. The error
    bars indicate the statistical uncertainties and the square
    brackets show the systematic errors. The STAR data points, for 0--5\%, 5--10\%, 10--20\%, 20--30\% and 30--80\% Au--Au collisions~\cite{Adams:2003ra}, are slightly shifted for clarity.
  }
\label{fig:relative_Radii}
\end{figure}
Figure~\ref{fig:relative_Radii}~\cite{Adamova:2017opl} shows the relative amplitudes of the
radius oscillations $R^{2}_{{\rm out},2}/R^{2}_{{\rm
    side},0}$,~$R^{2}_{{\rm side},2}/R^{2}_{{\rm side},0}$,
$R^{2}_{{\rm long},2}/R^{2}_{{\rm long},0}$, and $R^{2}_{{\rm
    os},2}/R^{2}_{{\rm side},0}$. We observe similar results to that of STAR, however the
    STAR results~\cite{Adams:2003ra,Adams:2004yc}
show on average larger oscillations for $\rside^{2}$. Our relative
amplitudes for $R^{2}_{\rm out,2}/R^{2}_{\rm side,0}$, $R^{2}_{\rm
  side,2}/R^{2}_{\rm side,0}$, and $R^{2}_{{\rm os},2}/R^{2}_{{\rm
    side},0}$ show a clear centrality dependence, whereas the $R^{2}_{\rm long,2}/R^{2}_{\rm long,0}$
is very close to zero for all centralities, similarly to the results
from RHIC~\cite{Adams:2003ra}.

\begin{figure}[!ht]
\centering
\includegraphics[width=8cm,keepaspectratio]
{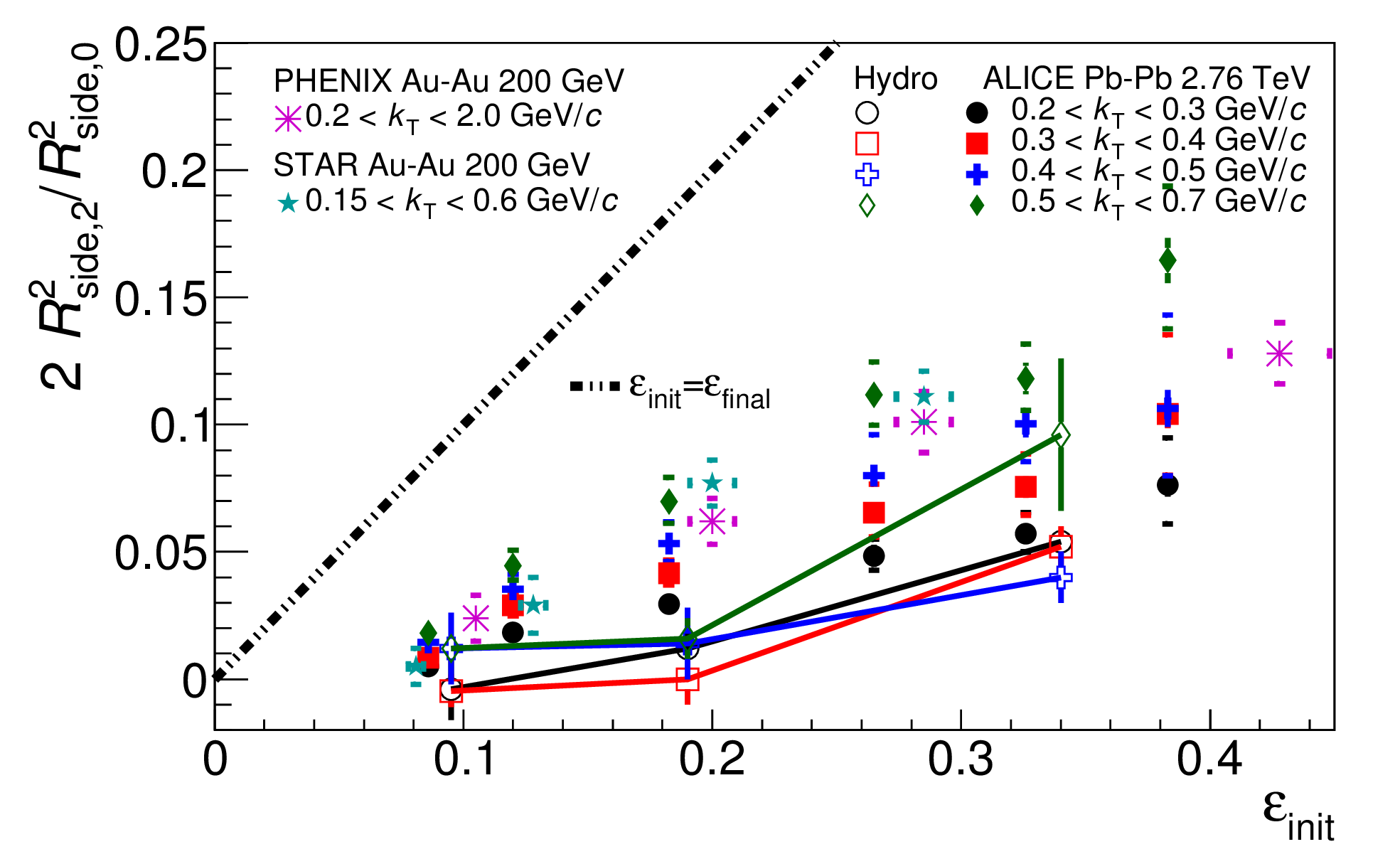}
  \caption{An estimate of freeze-out eccentricity $2
    R^{2}_{\rm side,2}/R^{2}_{\rm side,0}$ for different $\kt$ ranges vs.
    initial state eccentricity from Monte Carlo Glauber
    model~\cite{Ghosh:2016npo} for six centrality ranges, 0--5\%,
    5--10\%, 10--20\%, 20--30 \%, 30--40 \%, and 40--50\% from ALICE compared to RHIC results~\cite{Adare:2014vax,Adams:2003ra} and
    3+1D hydrodynamical calculations~\cite{Bozek:2014hwa}. Square brackets
    indicate systematic errors.}
\label{fig:finaleccentricity}
\end{figure}

The final source eccentricity at freeze-out
$\eps_{\rm final}$ can be estimated with an accuracy within 20--30\% as
$\eps_{\rm final}\approx
2\rsidetwo^2/\rsidez^2$~\cite{Retiere:2003kf}.
Figure~\ref{fig:finaleccentricity}~\cite{Adamova:2017opl} presents $2\rsidetwo^2/\rsidez^2$
for different $\kt$ ranges as a function of the initial-state
eccentricity for six different centralities.
We find a smaller final-state anisotropy in
the LHC regime compared to RHIC energies~\cite{Adare:2014vax,Adams:2003ra}. In Fig.~\ref{fig:finaleccentricity},
we also compare our results to the 3+1D hydrodynamic
calculations~\cite{Bozek:2014hwa}, this model
slightly underestimates the final source eccentricity.\\
\begin{figure}[!ht]
\centering
\includegraphics[width=11cm,keepaspectratio]
{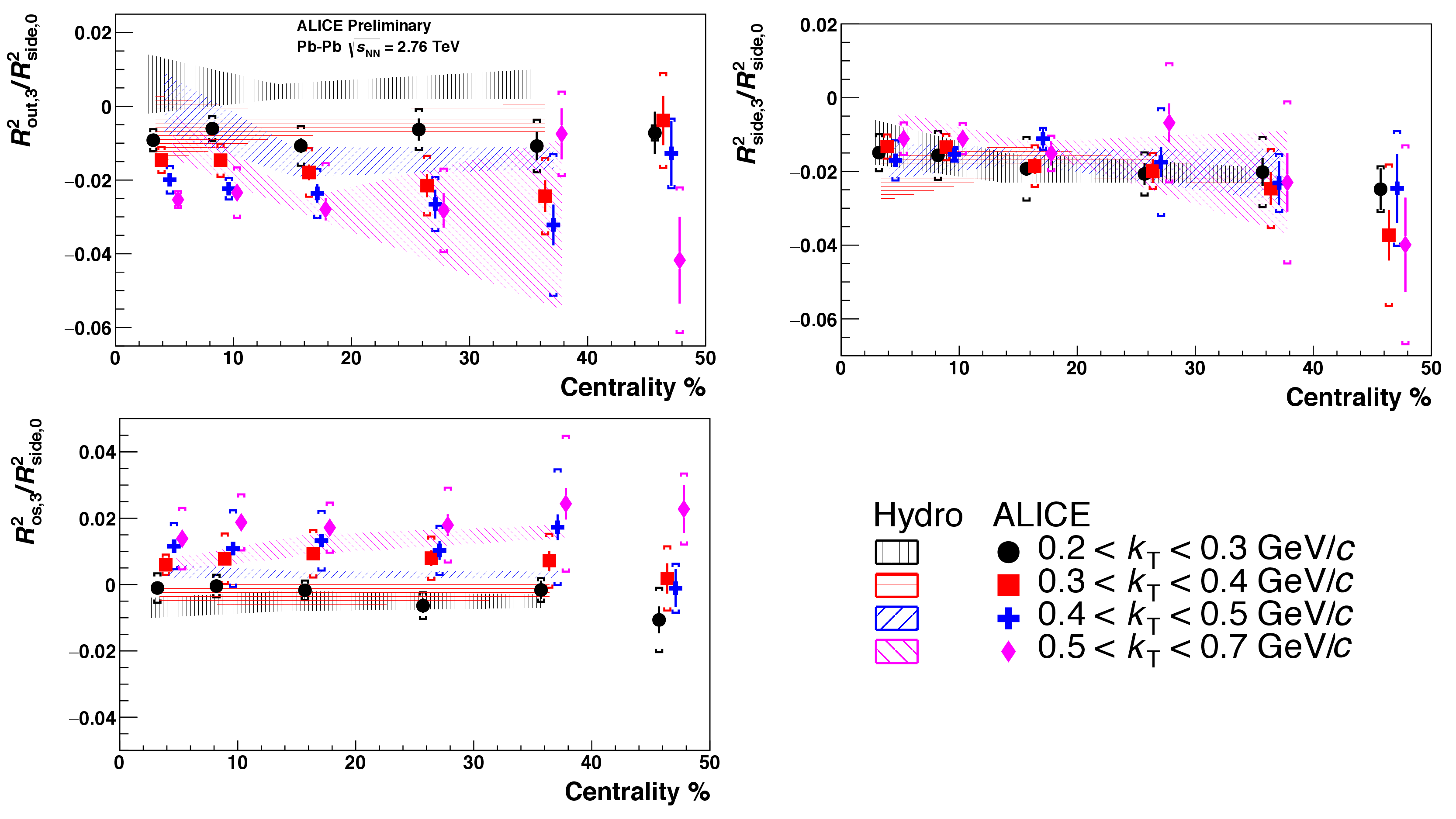}
  \caption{Amplitudes of the relative radii oscillations~$R^{2}_{\rm
      out,3}/R^{2}_{\rm side,0}$,~$R^{2}_{\rm side,3}/R^{2}_{\rm
      side,0}$, and~$R^{2}_{\rm os,2}/R^{2}_{\rm side,0}$ versus centrality for four $\kt$ ranges. The shaded bands are the 3+1D hydrodynamical calculations. Square brackets indicate systematic errors.
}
\label{fig:Third}
\end{figure}
\begin{figure}[!ht]
\centering
\includegraphics[width=14cm]
{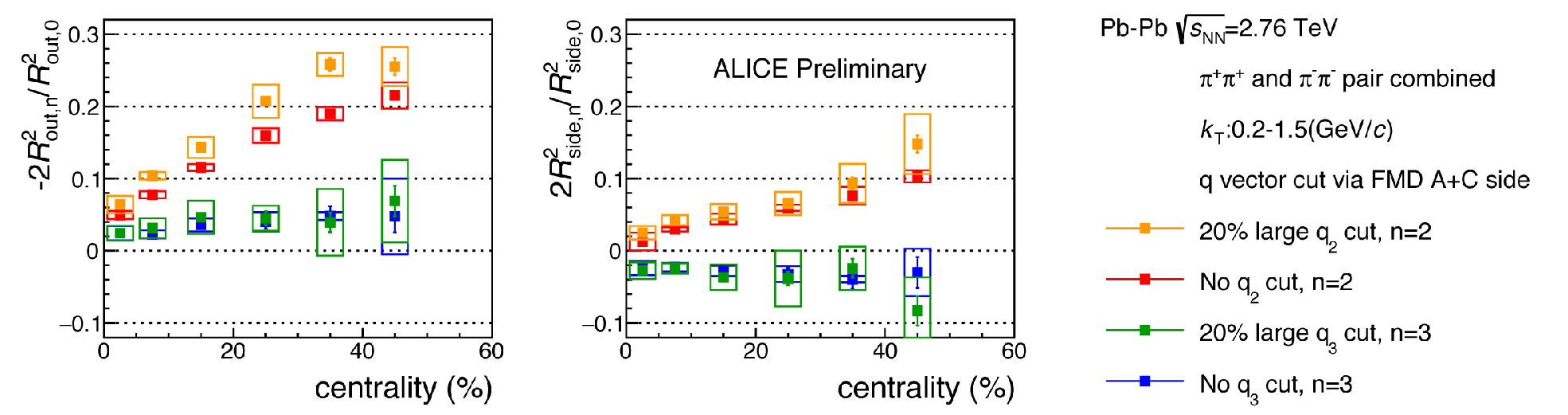}
  \caption{Amplitudes of the relative radius oscillations versus centrality with and without large $q_{2}$ ($q_{3}$) selection.}
\label{fig:ESE}
\end{figure}
Figure \ref{fig:Third} shows the relative amplitudes of the
radius oscillations $R^{2}_{{\rm out},3}/R^{2}_{{\rm
    side},0}$,~$R^{2}_{{\rm side},3}/R^{2}_{{\rm side},0}$, and $R^{2}_{{\rm
   os},3}/R^{2}_{{\rm side},0}$. The relative
    amplitudes for $R^{2}_{\rm out,3}/R^{2}_{\rm side,0}$ and $R^{2}_{\rm
      side,3}/R^{2}_{\rm side,0}$ have negative values for all centralities and $\kt$ cuts. We compare our results
       with the 3+1D hydrodynamic calculations~\cite{Bozek:2014hwa}, where the relative amplitudes $R^{2}_{{\rm side},3}/R^{2}_{{\rm side},0}$ and $R^{2}_{{\rm os},2}/R^{2}_{{\rm
    side},0}$ agree quantitatively, and the relative amplitudes $R^{2}_{{\rm out},3}/R^{2}_{{\rm
           side},0}$ agree qualitatively with the 3+1D hydrodynamical calculations~\cite{Bozek:2014hwa}. The relative amplitudes of the third harmonic results exhibit weak centrality and $\kt$ dependence. According
        to the 3+1D hydrodynamical calculations, the negative signs of oscillations of $R_{\mathrm out}$ and $R_{\mathrm side}$ are an indication that the initial triangularity
        has been washed-out or even reversed due to the triangular flow.
In the event shape engineering analysis, events were selected based on the
magnitude of the second (third) order flow vector $q_{2}$ ($q_{3}$)~\cite{Schukraft:2012ah}, the Forward Multiplicity detector (FMD) was used to select on the magnitude of the flow vectors (-3.4~$<$~$\eta_{\mathrm{FMDC}}$~$<$~-1.7, 1.7~$<$~$\eta_{\mathrm{FMDC}}$~$<$~5). We studied the effect of selecting the top 20\%
of the flow vector $q_{2}$ ($q_{3}$) on the magnitude of the flow $v_{2}$ ($v_{3}$). An enhancement of about 25\% (15\%) for $v_{2}$ ($v_{3}$) was observed
for all centralities. Figure \ref{fig:ESE} shows the effect of large $q_{2}$ ($q_{3}$) selection on the relative amplitudes of the radii oscillations $R^{2}_{{\rm out},2}/R^{2}_{{\rm
    side},0}$ ($R^{2}_{{\rm out},3}/R^{2}_{{\rm
        side},0}$) and~$R^{2}_{{\rm side},2}/R^{2}_{{\rm side},0}$ (~$R^{2}_{{\rm side},3}/R^{2}_{{\rm side},0}$). The large $q_{2}$ selection significantly enhances the
    relative amplitudes of the radius oscillations $R^{2}_{{\rm out},2}/R^{2}_{{\rm
        side},0}$ and slightly enhanced ~$R^{2}_{{\rm side},2}/R^{2}_{{\rm side},0}$, possibly selecting more elliptic initial source. However, the large $q_{3}$
        selection doesn't affect the relative amplitudes of the radius oscillations.
\section{Summary}
We have performed a measurement of two-pion azimuthally differential femtoscopy relative to the second and third harmonic flow plane in Pb--Pb collisions at $\snn$~=~2.76 TeV. The final-state
source eccentricity, estimated via side-radius oscillations relative to the second harmonic flow plane, is noticeably smaller than at lower collisions energies, but
still exhibits an out-of-plane elongated source at freeze-out even after a stronger in-plane expansion. The negative signs of the relative amplitudes of radii oscillation,
 $R_{\mathrm out}$ and $R_{\mathrm side}$, measured relative
to the third harmonic flow plane, are an indication that the initial triangularity is washed-out or even reversed according to the 3+1D hydrodynamic calculations. The azimuthally differential
HBT combined with ESE measurements were performed. The large $q_{2}$ selection has significant enhancement on the HBT radii relative to the second harmonic flow plane, whereas
large $q_{3}$ selection has no effect on the HBT radii relative to the third harmonic flow plane.
\section*{Acknowledgement}
This material is based upon work supported by the U.S. Department of
Energy Office of Science, Office of Nuclear Physics under Award
Number DE-FG02-92ER-40713.
\bibliographystyle{elsarticle-num}
\bibliography{reference}






\end{document}